# Do Temperate Rocky Planets Around M Dwarfs have an Atmosphere?

René Doyon
Université de Montréal, département de physique
Trottier Institute for research on Exoplanet and Observatoire du Mont Mégantic

Abstract

Detecting an atmosphere, as the first step in assessing the potential habitability of nearby temperate planets, is one of the most important scientific objectives of the Webb mission, an endeavour in practice limited to a handful of well-characterized planets: Trappist-1d, e, f, g, LHS1140b, and the mini-Neptune K2-18b. The first 18 months of atmospheric characterization with JWST have confirmed both its power and versatility to probe exoplanet atmospheres, and have highlighted the challenge that stellar activity could pose to studying those atmospheres through transmission spectroscopy. Assessing the prevalence of atmospheres in temperate planets with a minimal degree of confidence will require a multi-cycle program of order of a few 1000 hours involving both eclipse photometry and transmission spectroscopy, a program that would have to be executed over a significant fraction of JWST's lifetime. Implementing specialized fast readout modes tailored for time-series observations of relatively bright stars would greatly enhance the efficiency and productivity of JWST's exoplanet science programs. The forthcoming 500 hours of Cycle 3 Director Discretionary Time dedicated to exoplanet programs represent a unique opportunity to initiate a deep reconnaissance of habitability of the best keystone temperate planets.

## 1. Introduction

One of the most important scientific questions that the James Webb Space Telescope mission should answer is whether temperate rocky planets orbiting nearby low-mass M dwarfs possess an atmosphere and liquid water on their surface. On one hand, M dwarfs present ideal targets for this exploration due to their smaller masses and radii, which simplifies the task of characterizing the mass, radius, and atmosphere of their planets compared to Earth-like analogs orbiting solar-type stars. This is known as the "M-dwarf opportunity." However, M dwarfs are also known for their intense stellar activity, including strong XUV radiation, an important source of atmospheric erosion. It remains very speculative whether the secondary atmosphere of small rocky planets can survive for billions of years within such a hostile radiation environment (Dong et al., 2020; Kite & Barnett 2020). Addressing this question holds significant implications for planning future flagship space missions and ground-based giant observatories aimed at detecting life beyond the solar system.

This paper highlights the main results and lessons learned from the past 18 months of atmospheric characterizations of small planets orbiting M dwarfs with JWST. We discuss the roadmap needed to assess, with a minimal degree of confidence, the habitability of nearby temperate planets orbiting M dwarfs. We contend that given JWST's limited lifetime, a comprehensive program of both eclipse and transmission spectroscopy of a key sample of temperate planets, that is initiated as soon as possible, is an essential way forward.

## 2. The best temperate transiting planets: the Golden-J sample

The most suitable temperate planets for atmospheric characterization are those with equilibrium temperatures ($T_{eq}$) low enough to avoid runaway greenhouse atmospheric conditions that could turn water into a supercritical fluid state (Aguichine et al., 2021; Turbet 2023). While this depends on several atmospheric parameters and spectral type of the host star, we conservatively adopt $T_{eq}$ ~300 K as the maximum allowed value (assuming a Bond albedo $A_B$=0). These planets should also be required to have accurate mass and radius measurements to yield a planet bulk density ($\rho$) with an accuracy better than ~10%, which ultimately sets the precision on the mean molecular weight ($\mu$) inferred from transmission spectroscopic observations since the transit depth variation due to the atmosphere scales as $T_{eq} / \mu\rho R_*^2$. These criteria limit the selection to only a handful of rocky planets: Trappist-1d, e, f, g, and LHS1140b. We make an exception to include the temperate mini-Neptune K2-18b as a potential habitable world, despite its mass uncertainty of 18%. These planets, hereafter referred to the Golden-J sample[1], represent the temperate subset of the overall population of Earths/super-Earths and mini-Neptunes, all characterized by a bimodal radius distribution with a radius valley near 1.6-8 $R_\oplus$ (Fulton et al., 2017), with the Trappist-1 planets on the rocky side, LHS1140b in the middle of the valley, and K2-18b on the mini-Neptune side.

## 3. Highlights of the first atmospheric reconnaissance of small planets

Several important results and lessons learned have emerged from the first JWST observations both from emission and transmission spectroscopy program. Here we briefly highlight published results and ongoing analyses through private communications courtesy of several General Observations (GO) and Guaranteed Time Observations (GTO) teams.

3.1 The Trappist-1 planets
While the initial data suggests that TRAPPIST-1 b and c may be airless worlds (Greene et al., 2023; Zieba et al., 2023), the presence of atmospheres on these planets has not been definitively ruled out (Zieba et al., 2023; Ih et al., 2023; Lincowski et al., 2023). ***These first results nonetheless highlight the power of eclipse photometry to put strong constraints on the presence of an atmosphere on small rocky planets.***

Reconnaissance transmission spectroscopy of several planets (b, c, d, e, f, and g) show, in general, strong evidence of stellar activity (Lim et al., 2023) in the form of flares and unocculted spots and faculae that significantly bias the transmission spectra via the transit light source (TLS) effect. TLS (Rackham et al., 2018) is the stellar contamination coming from the assumption that the light from the planet's transit chord is the same as the disk-integrated light of the star when the planet is out of transit. If the star has unocculted spots/faculae then the assumption is invalid and induces stellar spectral signatures on the transit spectra of the planets, thereby complicating the interpretation of the detected spectral features. This fundamental problem is not specific to Trappist-1 and has been shown to affect many other transmission spectroscopy programs focussed on hot Jupiters even those with mid-K host stars (e.g. Fournier-Tondreau et al., 2024) thought to have relatively low level of stellar activity.

---

[1] In reference to a JWST version of the Golden sample, the handful nearest southern habitable worlds amenable to biosignature search with ANDES on the Extremely Large Telescope (Palle et al., 2023).

The strong level of stellar activity of Trappist-1, in particular the TLS effect, poses a significant challenge to unveil relatively small (~50 ppm) atmospheric planetary signals. Several lessons learned from stellar activity have emerged, namely: 1) the need for short-wavelength observations to lift the degeneracy between atmospheric and stellar activity signal (Moran et al., 2023). 2) Repeated observations should be favored to better unveil the inherent variability nature of stellar activity (Lim et al., 2023). 3) A comprehensive campaign of stellar activity characterization through spectroscopic monitoring of the Trappist-1 star over one full rotation is called for, to characterize the spectroscopic properties of flares and better understand stellar heterogeneities (de Wit et al., 2023, Berardo et al., 2024). 4) Finally, more specific to Trappist-1, a double-transit observing strategy should be favored consisting of observing a given transit event paired with another one from b under the assumption that Trappist-1b is airless. The latter spectrum can then be used, to first order, as a TLS "beacon" to correct the transmission spectrum of the planet of interest (de Wit et al., 2023).

3.2 K2-18b
A hydrogen-dominated atmosphere was unequivocally detected on the temperate mini-Neptune K2-18b using HST/WFC3 observations, showing a strong absorption feature at 1.4 $\mu$m (Benneke et al., 2019; Tsiaras et al., 2019) interpreted as water vapor. Benneke et al. argued that water can potentially condense into clouds in K2-18b atmosphere or even rain out. This discovery triggered the hypothesis that this planet could potentially be a Hycean world, a planet with a hydrogen atmosphere on top of a liquid water ocean (Madhusudan 2020, 2023a). Early JWST observations have provided a different picture showing a transmission spectrum dominated by methane (Madhusudhan et al., 2023) interpreted as evidence for the Hycean hypothesis which is challenged by other studies (e.g. Hamish et al., 2024; Wogan et al., 2024). While speculative, Hycean worlds may be a shortcut to biosignature detection with JWST, for instance through the dimethyl sulfide molecule featuring several molecular bands beyond ~3 $\mu$m. More work is needed to understand whether these planets are indeed habitable (Wogan et al., 2024) or if the DMS is more likely to be due to life or a planetary process. **K2-18b serves as a reminder that the search for life in the universe should not be limited to only rocky planets that match the characteristics of Earth.**

3.3 LHS1140b
This M4.5 dwarf is host of two small planets including LHS1140b which is the closest transiting temperate planet after those of the Trappist-1 system. The mass and radius and LHS1140b are now known with exquisite accuracy (a few %) similar to the Trappist-1 planets. LHS1140b's bulk density cannot be explained by an Earth-like composition (Cadieux et al., 2024). Detailed internal structure modelling, informed by stellar abundance measurements of refractory elements (Fe, Mg, Si), yields two likely scenarios for the nature of LHS1140b: a mini-Neptune with a small (~0.1%) H/He-rich envelope or a water-world with a water mass fraction of 14±5% (Cadieux et al., 2024). Recent transmission spectroscopy of LHS1140b (DD6543) with NIRISS SOSS of LHS1140b shows a rather flat spectrum with a clear TLS facula-dominated signal (Cadieux et al., in prep). The mini-Neptune scenario is excluded by these recent observations with high significance (See Figure 1).

Cadieux et al. (2024) presented detailed GCM models for the water world scenario suggesting that LHS1140b may be a snowball with a liquid patch at the sub-stellar point if it has an Earth-like level of $CO_2$ (400 ppm), or more, it its atmosphere. The transmission spectrum predicted by water world GCM models is heavily muted by $H_2O$/$CO_2$ clouds leaving only a small (~15-20 ppm) $CO_2$ features at 2.8 and 4.3 µm. **LHS1140b is arguably the best temperate planet from which liquid surface condition may be inferred indirectly through the detection of an appropriate level of $CO_2$ in its atmosphere.**

One important observational constraint of LHS1140b is its limited visibility. Only 8 visits (4 transits, 4 eclipses) are possible per year with JWST. As shown in Cadieux et al. (2024), it will likely take a dozen of visits (three years) to reach a minimal 3σ detection of $CO_2$ through transmission spectroscopy. Four visits (one year) with MIRI will be required to reach a 3σ detection of the secondary eclipse assuming the airless case with no heat redistribution (see Table 1).

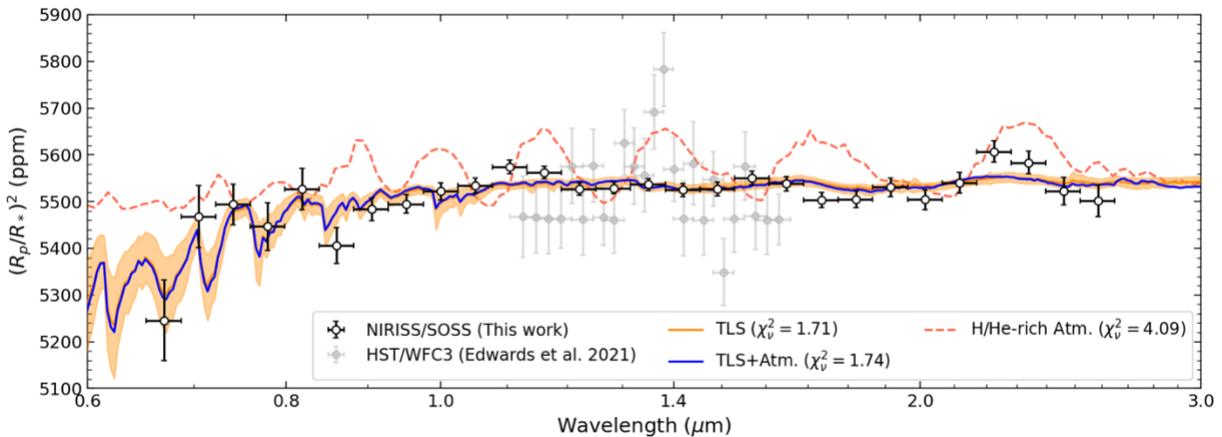

Figure 1 – Combined transmission spectrum LHS1140b obtained with NIRISS SOSS on December 1st and 26th 2023 overlaid with HST observations (solid grey points; Edwards et al. 2021). The orange 2σ-envelope is the best TLS model dominated by facula. The dashed line is the predicted GCM model of the mini-Neptune scenario of Cadieux et al. (2024); this specific model is excluded at more than10 σ (Cadieux et al., in prep).

## 4. For a deep habitability reconnaissance of the Golden-J planets

Arguably, the single most important question that JWST should and *can* answer, under the scientific theme *Planetary Systems and the Origin of Life,* is the title of this paper: Do temperate rocky planets around M dwarfs have an atmosphere? As mentioned above, the best characterized planets to answer this key question is the Golden-J sample. How much observing time would be required to reach an unambiguous answer for this sample? Given the huge implication of the outcome, any results should be independently confirmed with *both* eclipse and transmission observations with some minimal ($3\sigma$) level of confidence from both modes. One should indeed take advantage of the full power of JWST to answer such a fundamental question with far reaching implications. A habitability reconnaissance of the Golden-J planets (see Table 1) should have the following two minimal components for the rocky planets:

1) <u>15 $\mu$m eclipse photometry with MIRI:</u> goal to seek a 3-sigma detection of the airless planet scenario (no heat redistribution, $A_B$=0), i.e., the deepest possible eclipse depth.
2) <u>Deep reconnaissance transmission spectroscopy:</u> goal to seek a 3-sigma detection of the 4.3 $\mu$m $CO_2$ absorption feature.

Table 1 presents realistic estimates of the required observing time based on previous eclipse and transmission spectroscopy programs. This comprehensive reconnaissance effort would necessitate a minimum of ~700 hours, including approximately 200 hours dedicated to eclipse photometry. The latter could be conducted as early as Cycle 3 as part of the 500-hour Director Discretionary Time (DDT) program allocated to exoplanets. For the transmission spectroscopy component of the program, only Trappist-1e (128 hours) will be observed during the next two cycles (3 and 4) as part of GO programs. Given the current recommendation that the DDT program should solely focus on eclipse photometry, other equally important transmission spectroscopy programs on Trappist-1f, Trappist-1g, and LHS1140b would not commence until Cycle 4 through standard GO programs[2]. Imposing a higher detection threshold (4-5σ) for this reconnaissance program - as was published for the MIRI observations of Trappist-1b and c - would significantly increase the total observing time, potentially ranging from 1300 to 2000 hours, especially when considering the likelihood of extensive follow-up programs triggered by potential tantalizing detections on any of the targeted planets.

This exercise is intended to demonstrate that thoroughly assessing the prevalence of atmospheres in the most promising rocky temperate planets, with a high degree of confidence, will likely necessitate of order a few thousands of hours of observation, spanning a significant portion of JWST's operational lifetime and most likely extending beyond its mission Level 1 requirement of 5 years. Initiating such an extensive habitability program at the earliest opportunity is paramount. The upcoming 500-hour DDT program presents a unique chance to kickstart an in-depth habitability reconnaissance of the Golden-J sample, potentially marking a pivotal moment for JWST in addressing one of its most important scientific questions.

---

[2] LHS1140b will not be observed in cycle 3 as part of the normal GO program.

The success of exoplanet science inevitably requires a substantial portion of JWST's observing time to be dedicated to observing relatively bright stars. In this regard, it is crucial for the JWST Project to promptly implement specialized fast readout modes to increase the dynamic range of both the NIRSpec prism and NIRISS SOSS modes. Implementing these new readout modes would significantly enhance the Observatory's efficiency to enable more and better exoplanet science.

## Acknowledgement

We owe special thanks to Charles Cadieux, Olivia Lim, Vikki Meadows, Nikku Madhusudhan, Julien de Wit, Nikole Lewis, and all participants of the PAS workshop for their enlightening discussions on JWST science, society, and public outreach.

**Table 1 – JWST Habitability Reconnaissance Program of the Golden-J Planets**

| 15 μm Eclipse photometry (MIRI) | | | | | |
|---|---|---|---|---|---|
| **Planet** | **$T_{max}$ (K)[a]** | **$F_p/F_*$ (ppm)** | **$N_{visit}$[b]** | **$T_{exp}$ (hrs)[c]** | **Cycle** |
| Trappist-1d | 366 | 203 | 7 | 24 | 3 |
| Trappist-1e | 319 | 183 | 8 | 30 | 3 |
| Trappist-1f | 279 | 148 | 11 | 45 | 3 |
| Trappist-1g | 252 | 120 | 15 | 66 | 3 |
| LHS1140b | 288 | 77 | 4 | 30 | 3 |
| | | | Sub-total | 195 | |
| **Transmission spectroscopy** | | | | | |
| **Planet** | **Instrument mode** | | **$N_{visit}$** | **$T_{exp}$ (hrs)** | **Cycle** |
| Trappist-1e[d] | NIRSpec prism | | 15 | 128 | 3,4 |
| Trappist-1f[e] | NIRSpec prism | | 15 | 140 | 4,5 |
| Trappist-1g[e] | NIRSpec prism | | 15 | 145 | 4,5 |
| LHS1140b[f] | NIRISS SOSS+NIRSpec G395 | | 12 | 77 | 4,5,6 |
| | | | Sub-total | 490 | |
| | | | Grand total | 685 | |

[a] Maximum temperature assuming no heat redistribution to the night side with $A_B$=0.
[b] Number of visits needed to achieve a 3σ detection of the eclipse depth for the airless case.
[c] Total exposure time assuming the same experimental design of Zieba et al (2023) equivalent to 1 hr + 3x$t_{14}$ per visit where $t_{14}$ is the transit duration.
[d] Already programmed for Cycle 3 and 4 (PID6456).
[e] Assumes the same double-transit experimental design for Trappist-1e.
[f] Assumes the $CO_2$-dominated spectrum of the water world scenario of Cadieux et al (2024). Details of the experimental design to be reviewed/optimized after 4 visits when stellar activity (TLS) is better characterized.

Zieba, S. et al. 2023, *No thick carbon dioxide atmosphere on the rocky exoplanet TRAPPIST-1 c*, Nature, 620, 746